\documentclass[11pt]{article}

\newcommand{\PUBD}{Definitions}

\newcommand{\FGSD}{Plots}  

\usepackage{verbatim,vmargin,amsmath,calc,mathcomp}
\usepackage{epsfig}
\usepackage{graphicx}
\usepackage{afterpage}
\usepackage{psfrag}
\usepackage{url}

\newcommand{\ra}{\ensuremath{\rightarrow}}

\newcommand{\mumu}{\ensuremath{\mu^{+}\mu^{-}}}

\newcommand{\nbel}{\ensuremath{{\overline{\nu}}_{e}}}

\newcommand{\GeV}{\ensuremath{\mathrm{GeV}}}
\newcommand{\MeV}{\ensuremath{\mathrm{MeV}}}

\newcommand{\nut}   {\nu_\tau}

\newcommand{\en}{\ensuremath{e \nbel}}

\newcommand{\GeVm}{\ensuremath{\mathrm{GeV/c{^2}}}}
\newcommand{\GeVp}{\ensuremath{\mathrm{GeV/c}}}

\newcommand{\ifb}{\ensuremath{\mathrm{fb^{-1}}}}

\newcommand{\bfl}{\begin{flushleft}}
\newcommand{\efl}{\end{flushleft}}
\newcommand{\bfr}{\begin{flushright}}
\newcommand{\efr}{\end{flushright}}
\newcommand{\bc}{\begin{center}}
\newcommand{\ec}{\end{center}}

\newcommand{\Like}{\ensuremath{\mathcal{L}}}
\newcommand{\Acp}{\ensuremath{\mathcal{A}_{CP}}}

\newcommand{\pizero}  {\ensuremath{\pi^{0}}}

\newcommand{\pipm}    {\ensuremath{\pi^{\pm}}}
\newcommand{\Kpm}     {\ensuremath{K^{\pm}}}
\newcommand{\pimp}    {\ensuremath{\pi^{\mp}}}

\newcommand{\K}      {\ensuremath{K}}
\newcommand{\Kstar} {\ensuremath{\K^{*}}}
\newcommand{\Kstarz} {\ensuremath{\K^{*0}}}

\newcommand{\Kstarpm} {\ensuremath{K^{*\pm}}}

\newcommand{\rhoz}    {\ensuremath{\rho^{0}}}
\newcommand{\rhomp}   {\ensuremath{\rho^{\mp}}}
\newcommand{\rhopm}   {\ensuremath{\rho^{\pm}}}
\newcommand{\rhostar} {\ensuremath{\rho^{*}}}
\newcommand{\rhostarpm} {\ensuremath{\rho^{*\pm}}}

\newcommand{\Bpm}     {\ensuremath{B^{\pm}}}
\newcommand{\Bp}      {\ensuremath{B^{+}}}
\newcommand{\Bm}      {\ensuremath{B^{-}}}
\newcommand{\B}       {\ensuremath{B}}
\newcommand{\Bbar}    {\ensuremath{\bar{B}}}
\newcommand{\BpBm} {\ensuremath{B^{+}B^{-}}}

\newcommand{\BB}   {\ensuremath{B\bar{B}}}
\newcommand{\pipipi}  {\ensuremath{\pipm \pizero \pizero}}

\newcommand{\rhopi}   {\ensuremath{\rhopm \pizero}}

\newcommand{\Kstarpi} {\ensuremath{\Kstarpm \pizero}}

\newcommand{\Btopipipi}  {\ensuremath{\Bpm \ra \pipm \pizero \pizero}}
\newcommand{\Btorhopi}   {\ensuremath{\Bpm \ra \rhopi}}
\newcommand{\Btokstarpi} {\ensuremath{\Bpm \ra \Kstarpi}}
\newcommand{\Btorhostarpi} {\ensuremath{\Bpm \ra \rhostarpm\piz}}

\newcommand{\thetabsph} {\ensuremath{\theta_{\mathrm{Sph}}^{\mathrm{B}}}}  
\newcommand{\ANN}       {\ensuremath{{ANN}}}
\newcommand{\DeltaE}  {\ensuremath{\DeltaE}}
\newcommand{\mes}     {\ensuremath{m_{\mathrm{ES}}}}

\newcommand{\rhop} {\ensuremath{\rho^{+}}}
\newcommand{\rhom} {\ensuremath{\rho^{-}}}

\input{\PUBD/babarsym.tex}

\def\LUMI{211}  
\def\BBpairs{232}  
\def\BBpairsErr{3}  
\def\OffResLumi{22}  
\def\Yield{357} 
\def\YieldErr{49} 
\def\BRMean{10.0} 
\def\BRStat{1.4} 
\def\BRSyst{0.9} 
\def\AcpMean{-0.01} 
\def\AcpStat{0.13} 
\def\AcpSyst{0.02} 
\def\SigStat{8.7} 

\newcommand{\BABARPubYear}    {05}

\newcommand{\BABARConfNumber} {005}
\newcommand{\SLACPubNumber}  {11294}

\long\def\inst#1{\par\nobreak\kern 4pt\nobreak
    {\it #1}\par\vskip 10pt plus 3pt minus 3pt}

\begin{document}
{\pagestyle{empty}

\begin{flushright}
\babar-CONF-\BABARPubYear/\BABARConfNumber \\
SLAC-PUB-\SLACPubNumber \\
June 2005 \\
\end{flushright}

\par\vskip 5cm

\begin{center}
\Large \bf   Measurement of the {\boldmath{\Btorhopi\ }} Branching Fraction\\
  and Direct {\boldmath{\CP}} Asymmetry
\end{center}
\bigskip

\begin{center}
\large The \babar\ Collaboration\\
\mbox{ }\\
\today
\end{center}
\bigskip \bigskip

\begin{center}
\large \bf Abstract
\end{center}
An improved measurement of the process \Btorhopi\ is presented.
The data sample of \LUMI~\ifb\ comprises \BBpairs\ million \FourS\to\BB\ decays collected
with the \babar\ detector at the \pep2\ \abf\ at SLAC.
The yield and \CP asymmetry are calculated using an extended
maximum likelihood fitting method. The branching fraction and
asymmetry are found to be
$    \BR(\Btorhopi) = [\BRMean~ 
    \pm \BRStat~(Stat.)~ 
    \pm \BRSyst~(Syst.) 
    ] \times 10^{-6}$ 
    and
$    \Acp(\Btorhopi) = \AcpMean~ 
    \pm \AcpStat~(Stat.)~    
    \pm{\AcpSyst}~(Syst.)$, 
superseding  
previous measurements.
The statistical significance of the signal is calculated
to be \SigStat$\sigma$.
\vfill
\begin{center}
Contributed to the 
XXII$^{\rm st}$ International Symposium on Lepton and Photon Interactions at High~Energies, 6/30 --- 7/5/2005, Uppsala, Sweden
\end{center}

\vspace{1.0cm}
\begin{center}
{\em Stanford Linear Accelerator Center, Stanford University, 
Stanford, CA 94309} \\ \vspace{0.1cm}\hrule\vspace{0.1cm}
Work supported in part by Department of Energy contract DE-AC03-76SF00515.
\end{center}

\newpage
} 

\begin{center}
\small

The \babar\ Collaboration,
\bigskip

B.~Aubert,
R.~Barate,
D.~Boutigny,
F.~Couderc,
Y.~Karyotakis,
J.~P.~Lees,
V.~Poireau,
V.~Tisserand,
A.~Zghiche
\inst{Laboratoire de Physique des Particules, F-74941 Annecy-le-Vieux, France }
E.~Grauges
\inst{IFAE, Universitat Autonoma de Barcelona, E-08193 Bellaterra, Barcelona, Spain }
% ...

\end{center}\newpage

%
%
%
%
\section{Introduction}
\label{sec:intro}

Branching fraction and \CP asymmetry measurements of charmless \B meson decays
provide valuable constraints for the determination of the unitarity triangle
constructed from elements of the Cabibbo-Kobayashi-Maskawa quark-mixing
matrix~\cite{Cabibbo,KobayashiMaskawa}.
In particular, the angle
$\alpha \equiv \arg\left[-V_{td}^{}V_{tb}^{*}/V_{ud}^{}V_{ub}^{*}\right]$
of the unitarity triangle can be extracted from
decays of the \B meson to $\rho^\pm\pi^\mp$ final states~\cite{babarrhopi}.
However, the extraction is complicated
by the interference of decay amplitudes with differing weak
and strong phases.
One strategy to overcome this problem is to perform an SU(2) analysis
that uses all $\rho\pi$ final states~\cite{QuinnAndSnyder}. Assuming
isospin symmetry, the angle $\alpha$ can be determined free of
hadronic uncertainties from a pentagon relation formed in the complex plane
by the five decay amplitudes $B^0\to\rho^+\pi^-$,
$B^0\to\rho^-\pi^+$, $B^0\to\rho^0\pi^0$, $B^+\to\rho^+\pi^0$ and
$B^+\to\rho^0\pi^+$.
These amplitudes can be determined from measurements of the
corresponding decay rates and \CP asymmetries.
While all these modes have been measured, the current experimental uncertainties
need to be reduced substantially for a determination of $\alpha$.
Here we present an update to a previous measurement~\cite{babarrhopi0} of the
\Btorhopi\ branching fraction and \CP asymmetry
$$\Acp=\frac{N(\Bm\to\rhom\piz)-N(\Bp\to\rhop\piz)}{N(\Bm\to\rhom\piz)+N(\Bp\to\rhop\piz)}.$$

%
%
%
%
\section{Data Set and Candidate Selection}
\label{sec:cand_sel}

The data used in this analysis were collected with the \babar\ detector~\cite{BabarDet}
at the \pep2\
asymmetric-energy \epem\ storage ring
at SLAC.
Charged-particle trajectories are measured by
a five-layer double-sided silicon vertex tracker and
a 40-layer drift chamber
located within a 1.5-T solenoidal magnetic field.
Charged hadrons are identified by combining energy-loss
information from tracking with the measurements from 
a ring-imaging Cherenkov detector.
Photons are detected by
a CsI(Tl) crystal electromagnetic calorimeter
with an energy resolution of $\sigma_E/E=0.023(E/\GeV)^{-1/4}\oplus 0.014$.
The magnetic flux return is instrumented for muon and \KL\ identification.

The data sample
includes $\BBpairs\pm\BBpairsErr$~million \BB\ pairs
collected at the \FourS resonance,
corresponding to
an integrated
luminosity of \LUMI~\ifb.
It is assumed that
neutral and charged \B meson pairs
are produced in equal numbers~\cite{BBProdRatio}.
In addition, \OffResLumi~\ifb\ of data collected
40~\MeV\ below the \FourS resonance mass are used for background studies.

We perform full detector Monte Carlo (MC) simulations
equivalent to 
460 \ifb\ of generic \BB decays and 140 \ifb\ of continuum
quark-antiquark production events.
In addition, we simulate
over 50 exclusive charmless \B decay modes,
including 1.4 million signal \Btorhopi\ decays.

\B meson candidates are reconstructed from
one charged track and two neutral pions, with the following requirements:

\textbf{Track quality.}
The charged track used to form the \Btorhopi\ candidate is required to
have at least 12 hits in the drift chamber, to have a transverse momentum greater
than 0.1~\GeVp, and to be consistent with originating from
a \B-meson decay.
Its signal in the tracking and Cherenkov detectors is required to be
consistent with that of a pion.
We remove tracks that pass
electron selection criteria based on $dE/dx$ and calorimeter information.

\textbf{\piz\ quality.}
Neutral pion candidates are formed from two photons, each with a minimum energy
of 0.03~\GeV\ and a lateral moment~\cite{LATdef}
of their shower energy deposition
greater than zero and less than 0.6.
The angular acceptance of photons is restricted to exclude
parts of the calorimeter where showers are
not fully contained.
We require the photon clusters forming the \piz\ to be separated in space,
with a \piz\ energy of at least 0.2~\gev
and an invariant mass between 0.10 and 0.16~\GeVm.

\textbf{Kinematic requirements.}
Two kinematic variables,
$ \Delta E = E^{*}_{B} - \sqrt{s}/2$ and
the beam energy substituted mass of the \B-meson
$  \mes = \sqrt{(s/2 + {\bf p}_{0}\cdot {\bf p}_{B})^{2}/E^{2}_{0} - {\bf p}^{2}_{B}}$,
are used for the final selection of events.
Here $E^{*}_{B}$ is the energy of the $\B$ meson candidate
in the center-of-mass frame,
$E_{0}$ and $\sqrt{s}$ are the total energies of the $\epem$ system in
the laboratory and center-of-mass frames, respectively; ${\bf p}_{0}$ and
${\bf p}_{B}$ are the three-momenta of the $\epem$ system and the $\B$
candidate in the laboratory frame, respectively.
For correctly reconstructed \rhopi\ candidates \DeltaE\ peaks at zero, while
final states
with a charged kaon,
such as $\Btokstarpi$,
shift
$\DeltaE$
by approximately 80~\MeV\ on average.
Events are selected with
$5.20<\mes<5.29~\GeVm$  and
$|\DeltaE|<0.20~\GeV$.
The \DeltaE\ limits help remove background from two- and four-body \B decays
at a small cost to signal efficiency.

\textbf{Continuum suppression.}
Continuum quark-antiquark production is the dominant background.
To suppress it,
we select only those events where the angle \thetabsph\
in the center-of-mass frame
between the
direction of the \B meson candidate
and the
sphericity axis of the rest of the event
satisfies  $|\cos \thetabsph| < 0.9$.
In addition, we construct a non-linear
discriminant,  implemented as an artificial neural network, 
that uses three input parameters: 
the zeroth- and second-order
Legendre event shape polynomials $L_0,L_2$
calculated from the momenta and polar angles
of all charged particle and photon candidates
not associated with the \B meson candidate,
and the output of a
multivariate, non-linear \B meson candidate tagging algorithm~\cite{BTagRef}.
The output \ANN\ of the artificial neural network is peaked at 0.5
for continuum-like events and at 1.0 for \B decays.
We require $\ANN>0.63$ for our event selection.

\textbf{$\rho$ mass window.}
To further improve the signal-to-background ratio we
restrict
the  invariant mass of the $\rho$ candidate
to $0.55 < m_{\pi\pi} < 0.95~\GeVm$.

\textbf{Multiple candidates.}
Neutral pion combinatorics can lead to
more than
one \B-meson candidate per event.
We choose the best candidate based on a \chisq\ formed from
the measured masses of the two \pizero\ candidates within the
event compared to the known \pizero\ mass~\cite{pdg}.
In the case of multiple charged pion candidates the choice is random.

\textbf{Efficiency.}
The total \Btorhopi\ selection efficiency is $15.4\pm0.1\%$.
In MC studies, the signal candidate is correctly reconstructed 54.9\% of the time.
The remaining candidates come from
self-cross-feed (SCF, 37.5\%) and mistag events (7.6\%).
We define SCF events as those where one or more elements of the \B-candidate
reconstruction are incorrect except for its charge. They stem primarily
from swapping the low energy \piz from the resonance with another from the rest of the event.
Signal events reconstructed with the wrong charge are classified as mistag events.
Both SCF and mistag events emulate signal events, however the resolution in \mes\ and
\DeltaE\ tends to be worse.

%
%
%
%
\section{Background Contributions}
\label{sec:background}

MC events are used to study backgrounds from other \B-meson decays.
The dominant
contribution comes from $b \rightarrow c$ transitions;
the next most important is from
charmless \B-meson decays.
The latter tend to be more problematic as the branching fractions
are often poorly known, and because they may peak at the same
invariant mass as the
signal \Btorhopi\ events.
Seventeen individual charmless modes show a significant contribution
once
the event selection has been
applied (Table~\ref{tab:bbackmodes}).
These modes are added into the fit fixed at the yield and asymmetry
determined by the simulation, given an assumed branching fraction.
Wherever branching fractions are not available, we use half
the upper limit. If no charge asymmetry measurement is available,
we assume zero asymmetry.


\begin{table}
 \resizebox{1.0\textwidth}{!}{
 \renewcommand{\arraystretch}{1.3}
\begin{tabular}{|l||c|c|c|c|}
\hline \hline
\multicolumn{5}{|c|}{Backgrounds to \rhopi\ from $B$-related Sources}\\
\hline\hline
Mode & $\begin{array}{c} \mathrm{Efficiency} \\ (\%) \end{array}$ & 
$\begin{array}{c} \mathrm{Assumed}~ \mathcal{B} \\(\times 10^{-6})\end{array}$ 
& Assumed $\mathcal{A}_{CP}$ & $\begin{array}{c}\mathrm{Expected} \\ \mathrm{yield}\end{array}$\\
\hline
\hline
Generic $b\ra c$ & $1.69\times10^{-4}$      &      -    & 0.00     &   $392.3\pm19.8$ \\
$\Bz \ra \rhopm \rhomp$             & 2.12 & $30.0\pm6.0$ \cite{hfag}& $0.00\pm0.00$ & $147.7 \pm 29.5$\\
$\Bz \ra \rhopm \pimp$              & 2.39 & $24\pm2.5$ \cite{hfag}& $0.00\pm0.20$ & $132.9\pm13.8$ \\
$\Bpm \ra \pipm\piz$                & 4.12 & $5.5\pm0.6$ \cite{hfag}& $-0.02\pm0.07$\cite{hfag} & $52.5\pm5.7$\\
$\Bz \ra a_1^0\piz$                 & 1.16 & $17.5\pm17.5$ \cite{babarrhopi0} & $0.00\pm0.20$ & $46.9\pm46.9$ \\
$\Bpm \ra \rhopm \rhoz$             & 0.57 & $26.4\pm6.4$ \cite{hfag}& $-0.09\pm0.16$ \cite{hfag}& $34.6\pm8.4$ \\
$\Bpm \ra \pipm K_s(\ra\piz\piz)$   & 1.49 & $3.74 \pm0.20$    \cite{hfag}& $-0.02\pm0.03$\cite{hfag} & $12.9\pm0.7$ \\
$\Bpm \ra \Kpm\piz$                 & 0.41 & $12.1\pm0.8$ \cite{hfag}& $0.04\pm0.04$ \cite{hfag}& $11.5\pm0.8$ \\
$\Bz \ra \piz\piz$                  & 2.62 & $1.51\pm0.28$ \cite{hfag} & $0.00\pm0.00$ & $9.2\pm1.7$\\
$\Bz \ra \eta^\prime\piz$           & 1.15 & $1.85\pm1.85$    \cite{hfag}   & $0.00\pm0.20$ & $4.9\pm4.9$ \\
$\Bpm \ra \pipm \Kstarz$            & 0.17 & $9.76\pm1.22$  \cite{hfag}       & $0.00\pm0.20$ & $3.8\pm0.5$\\
$\Bz \ra \piz \Kstarz$              & 0.90 & $1.7\pm 0.8$ \cite{hfag}& $0.00\pm0.20$ & $3.5\pm1.7$\\
$\Bpm \ra \Kstarpi$                 & 0.61 & $2.3\pm0.8$ \cite{Kstarpi}& $0.04\pm0.29$ &  $3.4\pm1.0$\\
$\Bz \ra \Kpm \rhomp$               & 0.14 & $9.9\pm1.6$ \cite{hfag}& $0.17\pm0.16$\cite{hfag} & $3.2\pm0.5$ \\
$\Bpm \ra \Kstarpm \gamma$          & 0.02 & $40.3\pm2.6$ \cite{hfag}& $-0.01\pm0.03$\cite{hfag} & $2.1\pm0.1$ \\
$\Bpm \ra \rhopm \gamma$            & 0.65 & $0.9\pm0.9$ \cite{hfag}& $0.00\pm0.20$ & $1.4\pm1.4$ \\
$\Bz \ra \Kstarpm \rhomp$           & 0.05 & $12.0\pm12.0$ \cite{hfag}& $0.00\pm0.20$ & $1.4\pm1.4$ \\
$\Bpm \ra \rho^\pm(1450)\piz$       & -    & - & $0.00\pm0.20$ & $8\pm8$ \\
\hline
Total                               & -    & - & - & $872.2\pm62.1$\\
\hline
\hline
\end{tabular}
}
\caption[Summary of $B$ backgrounds to \Btorhopi.]{The
  individual \B--background modes considered in the fit.
  The expected number of events 
  after all cuts
  are listed for each mode.
  For modes which
  do not have well measured branching fractions, half the upper limit is used.
  If no \Acp\ measurement is available, we assume zero asymmetry with a 20\% uncertainty.}
\label{tab:bbackmodes}
\end{table}


\textbf{\rhostar\ resonances.}
Although all other states which decay like the $\rho$ to $\pi\piz$
-- subsequently referred to as \rhostar\ --
lie outside our $\rho(770)$ mass cut, a contribution to our
signal cannot be \emph{a priori} ruled out.
The only non-strange vector resonances which can decay to two
pions are the $\rho(1450)$ and the $\rho(1700)$.
To account for the possible presence of these modes, a fit to the
\Btorhostarpi\ yield is performed in a sideband of the invariant mass
using the three variables \mes, \DeltaE, and \ANN. The mass window
is chosen to be as far as possible from the $\rho(770)$ mass,
centered near the pole of the $\rho(1700)$
at $1.5 < m_{\pi\pi} < 2.0$~\gevcc. 
The fitted yield for the $B^\pm\ra\rho^{*\pm}\piz$ decay is then
extrapolated into the nominal region. Although
the choice of mass range is motivated by the $\rho(1700)$, any yield seen is
attributed entirely to the $\rho(1450)$, which is the closer of the two resonances
to the signal.
From the $B^\pm\ra\rho^\pm(1450)\piz$ MC, the ratio of
candidates in the sideband to candidates in the signal mass region is
approximately 12.6:1. The fit in the sideband yields $101 \pm 32$ events,
resulting in an estimate of the $\rho^*$ background of $8$ events.
We assign a conservative systematic uncertainty of 100\% to this number.
The $\rho^*$ then enters into the
nominal fit with PDFs constructed from $\B^\pm\ra\rho^\pm(1450)\piz$ MC simulation.

\textbf{Non-resonant decays to \pipipi.}
The non-resonant \Btopipipi\ branching fraction has, to date, not been
measured.
To estimate the significance of its
contribution we select a region of the Dalitz plot 
--- defined by the triangle $(m^2_{\pipm\piz_1},m^2_{\pipm\piz_2})=(6,6),(6,15),(11,11)$~GeV$^2/c^4$ ---
that is far
from the signal as well as $\rho(1450)$ and higher resonances and
which has low levels of continuum background. 
A likelihood fit in this region yields $-5.1\pm7.6$ non-resonant events
in a data sample of 1100 events. This is consistent with zero.
The non-resonant contribution is therefore deemed negligible.

%
%
%
%
\section{The Maximum Likelihood Fit}
\label{sec:thefit}

An unbinned maximum likelihood fit to the variables
\mes, \DeltaE, $m_{\pi\pi}$, and \ANN\
is used to extract
the total number of signal \Btorhopi\ and continuum background
events and their respective charge asymmetries.
The likelihood for the selected sample is given by the product
of the probability density functions (PDF) for each individual candidate,
multiplied by the Poisson factor:
$$    \Like = \frac{1}{N!}\,e^{-N^\prime}\,(N^\prime)^N\,\prod_{i=1}^N {\cal P}_i\, ,$$
where $N$ and $N^\prime$ are the number of observed and expected events, respectively.
The PDF ${\cal P}_i$ 
for a given event
$i$ is the sum of the signal and background terms:
\begin{eqnarray}
  {\cal P}_i
  & = &
  N^{\rm Sig} \times  \frac{1}{2} \,
  [\,(1-Q_i A^{\rm Sig}) f_{\rm Sig}\, {\cal P}^{\rm Sig}_{i} 
   \nonumber
  \\
  &  &
    ~+~ (1 - Q_i A^{\rm Sig})\, f_{\rm SCF}\, {\cal P}^{\rm Sig}_{{\rm SCF},i}     
  \nonumber
  \\
  &  &
    ~+~ (1+Q_i A^{\rm Sig}) f_{\rm Mis}\, {\cal P}^{\rm Sig}_{{\rm Mis},i} 
  \,]
  \nonumber
  \\
  & & 
  + \sum _j N^{\rm Bkg}_{j}\times  \frac{1}{2}  (1 - Q_iA^{\rm Bkg}_{j})\,  {\cal
    P}^{\rm Bkg}_{j,i},
  \nonumber 
  \label{pdfsum}
\end{eqnarray}
where $Q_i$ is the charge of the pion in the event,
$N^{\rm Sig}(N^{\rm Bkg}_j)$ and $A^{\rm Sig}(A^{\rm Bkg}_j)$
are the yield and asymmetry for
signal and background component $j$, respectively.
The fractions of true signal ($f_{\rm Sig}$), SCF signal ($f_{\rm SCF}$),
and wrong-charge mistag events ($f_{\rm Mis}$) are fixed to the numbers
obtained from MC simulations (Section~\ref{sec:cand_sel}).
The $j$ individual background terms comprise continuum, $b\to c$ decays, \rhostar,  and
17 exclusive charmless \B decay modes.
The PDF for
each component, in turn, is the product of the PDFs for each of the
fit input variables,
$  {\cal P} = {\cal P}_{{\rm m_{ES}},\Delta E}{\cal P}_{\ANN}{\cal P}_{m_{\pi\pi}}.$
Due to correlations between \DeltaE\ and \mes,
the ${\cal P}_{{\rm m_{ES}},\Delta E}$  for signal
and all background from \B decays are described by two-dimensional
non-parametric PDFs~\cite{2dkeys} obtained from MC events.
For continuum background,
${\cal P}_{{\rm m_{ES}},\Delta E}$ is the product of two orthogonal
one-dimensional parametric PDFs;
\mes\ is well described by
an empirical phase-space threshold
function~\cite{argus} and
\DeltaE\ is parameterized with a second degree polynomial.
The parameters of the continuum PDFs are floated in the fit,
with \mes\ constrained to masses below 5.29~\GeVm.
$\ANN$ is described by the product of an exponential and a
polynomial function for continuum
background and by a Crystal Ball function~\cite{CrystalBall} for all other modes.
For ${\cal P}_{m_{\pi\pi}}$, one-dimensional
non-parametric PDFs obtained from MC events
are used to describe all
modes except the signal mode itself, which is described by
a non-relativistic Breit-Wigner line-shape.
The parameters for this PDF are held fixed to the MC values and varied within
errors to estimate systematic uncertainties.

\begin{figure}[htbp]
  \begin{center}
    \epsfig{file=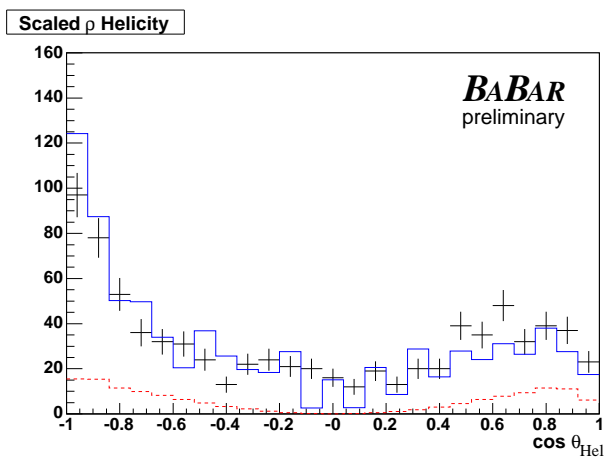,height=6cm}
    \caption{Distribution of $\cos\theta_{\rm Hel}$ for  $\mes>5.265$, $|\DeltaE|<0.1$, and $\ANN>0.85$.
      Data are shown in black, with error bars.
      The total PDF and the \Btorhopi\ contribution are overlaid in solid blue and dashed red lines, respectively.}
    \label{fig:helicity}
  \end{center}
\end{figure}

A number of cross checks confirm that the fit is unbiased.
In 1000 separate MC pseudo-experiments
we generate the expected number of events for the various fit
components before using the maximum likelihood fit to extract the yields and
asymmetries.
The distributions for each component are generated from the component's PDF,
giving values for the fit variables \mes, \DeltaE, \ANN, and $m_{\pi\pi}$.
The expected number of events is calculated from the branching fraction and
efficiency for each individual mode.
The generated number of events for each fit component is determined by fluctuating
the expected number according to a Poisson distribution.
The test is repeated using
samples with differing asymmetry values.
We repeat these MC studies
using
fully simulated
signal \Btorhopi\
events 
instead of generating the signal component from our PDFs.
This verifies that the signal component is correctly modeled
including correlations between the fit variables.
As another cross check we compare the distribution of the helicity angle
$\theta_{\rm Hel}$ between the momenta of the 
charged pion and the \B-meson in the $\rho$ rest frame
in data with that modeled in MC samples for a variety of cuts.
Fig.~\ref{fig:helicity} shows
the distribution of $\cos\theta_{\rm Hel}$
for a pseudo-signal-box defined by $\mes>5.265$,
$|\DeltaE|<0.1$, and $\ANN>0.8$. 
We generally find our PDFs in good agreement with the data.
Finally, omitting $m_{\pi\pi}$ as a fit variable
has no significant influence on the signal yield,
indicating that our treatment of \rhostar\ background
is indeed effective.

%
%
%
\section{Systematic Uncertainties}
\label{sec:systematics}

\renewcommand\baselinestretch{1.3}
\begin{table}[hbt]
  \begin{center}
  \caption[Systematic uncertainties]{Breakdown of systematic uncertainties.}
    \begin{tabular}{|ll|}
      \hline\hline      
      \multicolumn{2}{|c|}{Absolute uncertainties on yields}\\ 
      Source                   &  $\sigma_{\rm Syst.}^{\rm Yield}$ (Events)\\
      \hline
      Background normalization & $^{+14.1}_{-13.4}$ \\ 
      PDF shapes               & $^{+\;\, 4.7}_{-\;\, 4.2}$ \\ 
      SCF fraction             &  $\pm  12.2$ \\ 
      Mistag fraction          &  $\pm\;\,  2.0$ \\ 
      $\Delta$E shift          &  $\pm\;\, 2.6$  \\ 
      \hline
      Total                    & $\pm 19$ \\ 
      \hline\hline
      \multicolumn{2}{|c|}{Relative uncertainties on \BR(\Btorhopi) }\\
      Source            &  $\sigma_{\rm Syst.}^{\BR} (\%) $ \\
      \hline
      Efficiency estimation        &  $\pm 7.3 $ \\ 
      \B\ counting                 &  $\pm 1.1 $ \\
      \hline
      Total      & $\pm 7.4$ \\
      \hline\hline
      \multicolumn{2}{|c|}{Uncertainties on \Acp }\\
      Source            &  $\sigma_{\rm Syst.}^{\Acp }$ \\
      \hline
      Background normalization      &  $\pm 0.006$ \\ 
      Background asymmetry          &  $\pm 0.024$ \\ 
      PDF shapes                    &  $\pm 0.001$ \\
      \hline
      Total                         &  $\pm 0.02$  \\
      \hline\hline
    \end{tabular}
  \label{tab:systematics}
  \end{center}
\end{table}
\renewcommand\baselinestretch{1.0}

Individual contributions to the systematic uncertainty are
summarized in Table \ref{tab:systematics}.

\textbf{Absolute uncertainties on yields.}
\label{sec:syst_yields}
We calculate the uncertainty of the continuum background estimation
directly from the fit to data.
The
backgrounds from \B decays are determined from simulation and fixed
according to
their efficiencies and branching fractions.
The largest individual contribution comes from the $\B\to a_1^0\piz$ channel.
For those individual decay modes which have been measured,
we vary the number of events in the fit by 
their measured uncertainty.
For all others we vary the amount included in
the fit by $\pm 100\%$. For the
$b\to c$
component, we fix the rate
based on the number calculated from MC samples and vary the
amount based on the statistical uncertainty of this number.
The shifts in the fitted yields are calculated
for each mode in turn and then added in quadrature to find the total
systematic effect. The largest individual contribution comes from the
\rhostar\ estimation.

To take into account the variation of the two-dimensional
non-parametric PDFs used for
\DeltaE and \mes,
we smear the MC-generated distributions from which
the PDFs are derived.
This is effectively done by varying the kernel bandwidth~\cite{2dkeys} up
to twice its original value.
For
$m_{\pi\pi}$ and $\ANN$,
the parameterizations determined
from fits to MC events are varied by one standard deviation. 
The systematic uncertainties are determined using the altered PDFs and
fitting to the final data sample. The overall shifts in the central
value are taken as the size of the systematic uncertainty.

We vary the SCF fraction by
a conservative estimate of its relative uncertainty
($\pm10\%$) and assign the shift
in the fitted number of signal events as the systematic uncertainty of the SCF
fraction.

To account for differences in the neutral particle reconstruction
between data and MC simulation,
the signal PDF distribution in $\Delta$E is offset by $\pm 5 \mev$ and the data is then refitted. 
The larger of the two shifts in the central value of the yield is 2.2 events, which is taken
as the systematic uncertainty for this effect.

\textbf{Relative uncertainties on the branching fraction.}
\label{sec:syst_BR}
Corrections to the \pizero\ energy resolution and efficiency,
determined using various control samples,
add a systematic uncertainty
of 7.2\%.
A relative systematic uncertainty of $1\%$ is assumed
for the
pion identification.  
A relative systematic uncertainty of 0.8\% on the efficiency for a single charged
track is applied.
Adding all the above contributions in quadrature gives a
relative systematic uncertainty on the branching fraction of 7.3\%.
Another contribution of $1.1\%$ comes from
the uncertainty on the
total number of \B events.

\textbf{Uncertainties on the charge asymmetry.}
\label{sec:syst_Acp}
To calculate the effects of systematic
shifts in the charge asymmetries of background modes, each
mode is varied by its measured uncertainty.
For contributions with no measurement, we assume zero asymmetry and
assign an uncertainty of 20\%, motivated by the largest charge asymmetry
measured in any mode so far~\cite{Bkpi2004}.
The individual shifts are then added in
quadrature to find the total systematic uncertainty.
In addition, the effect
of altering the normalizations of the \B backgrounds affects the
fitted asymmetry.
The size of the shift on the fitted \Acp\ is taken as the size of
the systematic uncertainty.
%

%
%
\begin{figure}[p!]
  \renewcommand\baselinestretch{0.5}
  \begin{tabular}{cc}
    \epsfig{file=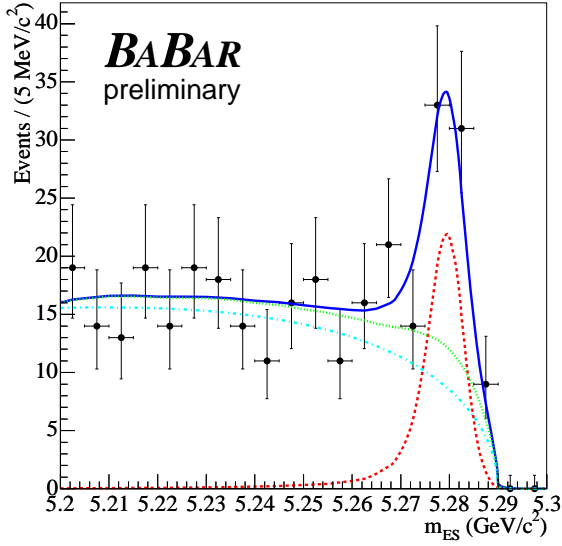,width=0.49\columnwidth}
    &
    \epsfig{file=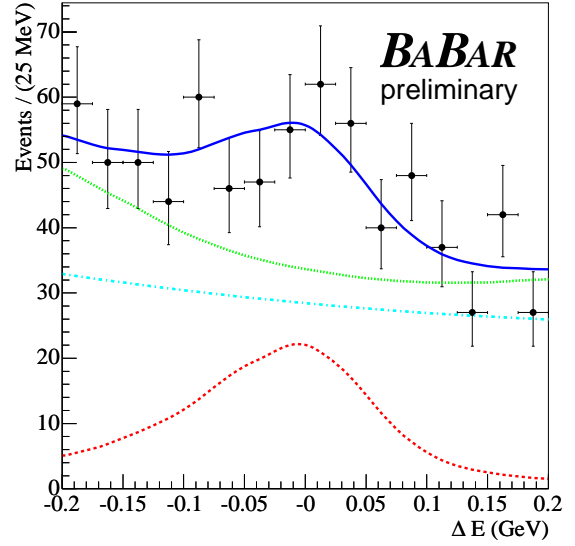,width=0.49\columnwidth}
    \\
    (a) & (b)\\
    \epsfig{file=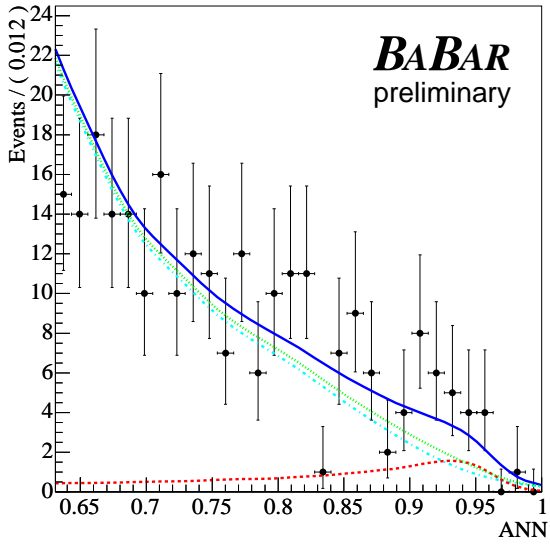,width=0.49\columnwidth}
    &
    \epsfig{file=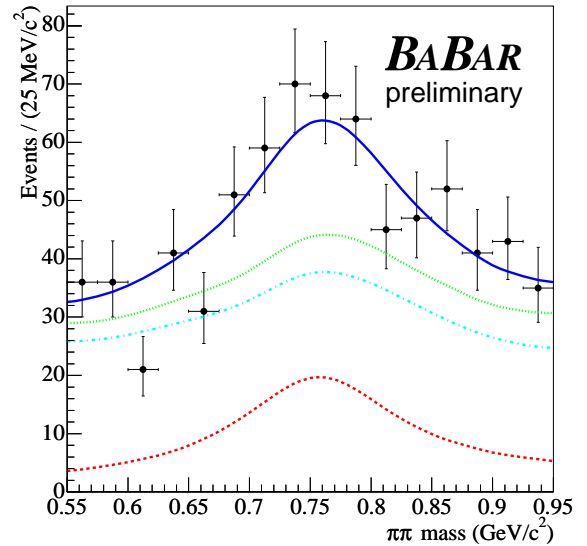,width=0.49\columnwidth}
    \\
    (c) & (d)\\
  \end{tabular}
  \renewcommand\baselinestretch{1.0}
  \caption[Results for the four variables of the maximum likelihood fit]
  {Likelihood projection plots for the four fit variables,
    (a) \mes, (b) \DeltaE, (c) \ANN, and (d) $m_{\pi\pi}$.
    In each plot the solid blue line represents the total PDF,
    the dotted green line represents the total background,
    the dotted-dashed light blue line represents the continuum contribution,
    and the dashed red line represents the signal component.
  The plots contain a subset of the events defined by a likelihood ratio
  of at least 0.1 (0.05 for \ANN).}
  \label{fig:results}
\end{figure}
%

%
%
\section{Results}
\label{sec:results}

The central value of the signal yield from the maximum likelihood fit is
$\Yield\pm\YieldErr$ events, over $44840\pm217$ continuum events and
an expected background of $872\pm 62$ events from other \B decays.
We find a branching fraction and charge asymmetry of
\begin{displaymath}
  \renewcommand{\arraystretch}{1.5}
  \begin{array}{ll}
    \BR(\Btorhopi) & = [\BRMean 
    \pm{\BRStat}~(Stat.)~ 
    \pm{\BRSyst}~(Syst.) 
    ] \times 10^{-6} \\
    \Acp(\Btorhopi) & = \AcpMean 
    \pm{\AcpStat}~(Stat.)~  
    \pm{\AcpSyst}~(Syst.). \\ 
  \end{array}
\end{displaymath}
Compared against the null hypothesis, the statistical significance
$\sqrt{-2 \ln (\Like_{Null}/\Like_{Max})}$ 
of the yield amounts to \SigStat\ standard deviations.
 
The results of the fit are illustrated in 
Fig.~\ref{fig:results}.
The plots are enhanced in signal
by selecting only those events which exceed a threshold of 0.1 (0.05 for \ANN)
for the likelihood ratio
$R = (N^{\rm Sig}{\cal P}^{\rm Sig})/(N^{\rm Sig}{\cal P}^{\rm Sig} + \sum_{i} N^{\rm Bkg}_{i}{\cal P}^{\rm Bkg}_{i})$,
where $N$ are the central values of the yields from the fit and
${\cal P}$ are the PDFs with the projected variable integrated out.
This threshold is optimized by maximizing the ratio
$S = N^{\rm Sig}~\epsilon^{\rm Sig}/\sqrt{N^{\rm Sig}~\epsilon^{\rm Sig} + \sum_{i} N^{\rm Bkg}_{i}~\epsilon^{\rm Bkg}_{i}}$
where $\epsilon$ are the efficiencies after the threshold is applied.
The PDF components are then
scaled by the appropriate $\epsilon$.

\section{Conclusions}
We have measured
the branching fraction and charge asymmetry for the
decay \Btorhopi\ using
a maximum likelihood fit. We obtain 
$\BR(\Btorhopi) = [\BRMean \pm \BRStat \pm \BRSyst]\times 10^{-6}$, and
$ \Acp = \AcpMean \pm {\AcpStat} \pm {\AcpSyst}$, respectively,
where the first error is statistical and the second error systematic.
The statistical significance of the signal
is calculated to be \SigStat\ standard deviations. The results are in good
agreement with the previous measurement~\cite{babarrhopi0}.

\section*{Acknowledgements}
We are grateful for the 
extraordinary contributions of our \pep2\ colleagues in
achieving the excellent luminosity and machine conditions
that have made this work possible.
The success of this project also relies critically on the 
expertise and dedication of the computing organizations that 
support \babar.
The collaborating institutions wish to thank 
SLAC for its support and the kind hospitality extended to them. 
This work is supported by the
US Department of Energy
and National Science Foundation, the
Natural Sciences and Engineering Research Council (Canada),
Institute of High Energy Physics (China), the
Commissariat \`a l'Energie Atomique and
Institut National de Physique Nucl\'eaire et de Physique des Particules
(France), the
Bundesministerium f\"ur Bildung und Forschung and
Deutsche Forschungsgemeinschaft
(Germany), the
Istituto Nazionale di Fisica Nucleare (Italy),
the Foundation for Fundamental Research on Matter (The Netherlands),
the Research Council of Norway, the
Ministry of Science and Technology of the Russian Federation, and the
Particle Physics and Astronomy Research Council (United Kingdom). 
Individuals have received support from 
CONACyT (Mexico),
the A. P. Sloan Foundation, 
the Research Corporation,
and the Alexander von Humboldt Foundation.

\bibliography{babar-conf-05005}
\bibliographystyle{h-physrev4}

\end{document}